# Mechanical behavior, enhanced dc resistivity, energy band gap and high temperature magnetic properties of Y-substituted Mg-Zn ferrites

M. A. Ali[1,*], M. N. I. Khan[2], M. M. Hossain[1], F.-U.-Z. Chowdhury[1], M. N. Hossain[3], R. Rashid[2], M. A. Hakim[3],S. M. Hoque[2],and M. M. Uddin[1,*]

[1]Department of Physics, Chittagong University of Engineering and Technology (CUET), Chattogram4349,Bangladesh.
[2]Materials Science Division, Atomic Energy Center, Dhaka 1000, Bangladesh.
[3]Department of Glass and Ceramic Engineering, Bangladesh University of Engineering and Technology (BUET), Dhaka 1000, Bangladesh.

## ABSTRACT

We report the synthesis of Y-substituted Mg-Zn [$Mg_{0.5}Zn_{0.5}Y_xFe_{2-x}O_4$ ($0 \leq x \leq 0.05$)] ferrites using conventional standard ceramic technique. The samples were characterized by X-ray diffraction (XRD) analysis, field emission scanning electron microscopy (FESEM), FTIR spectroscopy, UV–Vis spectroscopy and quantum design physical properties measurement system (PPMS).XRD patterns confirm the single phase cubic spinel structure up to $x = 0.03$ and appearance of a secondary phase of $YFeO_3$for higher Y contents. FESEM images depict the distribution of grains and EDS spectra confirmed the absence of any unwanted element. Completion of solid state reaction and formation of spinel structure has been revealed from FTIR spectra. The FTIR data along with lattice constant, bulk density and porosity were further used to calculate the stiffness constant ($C_{ij}$), elastic constant and Debye temperatures. Mechanical stability of all studied compositions is confirmed from $C_{ij}$ using Born stability conditions. Brittleness and isotropic nature are also confirmed using Poisson's ratio and anisotropy constants, respectively. The enhancement of dc electrical resistivity ($10^5\Omega$-cm to $10^6\Omega$-cm) with Y content is observed. The energy band gap (increased with Y contents) is found in good agreement with dc electrical resistivity. Ferrimagnetic to paramagnetic phase change has been observed from the field dependent high temperature magnetization curves. The magnetic moments and saturation magnetization were found to be decreased with increasing temperature. The Curie temperature ($T_c$) has been measured from temperature dependent magnetic moment (M-T) and initial permeability ($\mu'_i$-T) and found to be in good agreement with each other. Decrease in $T_c$ with Y content is due to redistribution of cations and weakening of the exchange coupling constant. The magnetic phase transition has been analyzed by Arrott plot and found to have second order phase transition. The dc resistivity endorses the prepared ferrites are suitable for high frequency and high temperature magnetic device applications as well.

## 1. Introduction

Since last century, the spinel type ferrites have come out as one of the leading materials in the field of materials science due to their vast applications. The physics involved in spinel structure is still interesting as their properties can easily be tuned, consequently their applications as well. In recent years, new opportunities have been disclosed and numerous directions have been developed for research on these materials by manipulating their physical properties. The long-established applications of the ferrites (in thin film, bulk or nano form) are in transformer applications, solar hydrogen production, multi-layer chip inductors, high density magnetic recording, telecommunications, microwave devices, as contrast agents in magnetic resonance imaging (MRI), sensors, catalysts etc. [1-8]. Moreover, the new application of spinel ferrites has been listed as photocatalysts for photocatalytic degradation of dyes [9-12].

Mg-Zn ferrite is one of the most used soft ferrites because of its high electrical resistivity, low cost, low dielectric loss, high mechanical hardness and superior environmental stability [13]. Mg-Zn ferrites with higher resistivity ($10^6$–$10^7$ Ω-cm) make its suitability in high frequency applications [13]. Most of (~90%) $Mg^{2+}$ ions are located on the *B*-sites and small fraction (~10%) occupies in the *A*-sites [14, 15]. However, the $Zn^{2+}$ ions have a preference to occupy *A*-sites in the spinel lattice [16, 17]. The $Fe^{3+}$ ions are distributed on both *A*- and *B*-sites [18]. Therefore, the physical properties of the Mg-Zn ferrites are determined by the cations distribution over the *A*- and *B*-sites. The physical properties can be altered by introducing the different metallic ions results the cations distribution modification on the *A*- and *B*-sites. The factors that determine cation distributions are ionic radius, charge, site preference and level of substituent ion, methods, conditions and sintering temperature for the preparation of ferrites [19-22].

Study of ferrites usually in most cases has been limited to their structural, electrical and magnetic properties [9-25]. However, study of mechanical properties of ferrites has scientific importance due to their potential applications in industry and in research field. Elastic constants are closely related to many important physical properties of solids, for instance acoustic-phonon vibration and internal stress, Debye temperature. This type of study provides the information about bonding nature within the materials. Also to understand the thermal properties and minimum changes in devices performance with mechanical shock, temperature, vibration, etc. during the manufacturing and testing of ferrite materials [26] Due to the mentioned interest the study of mechanical properties have already been considered by researchers [26-29]. Moreover, the study of magnetic properties of Mg-Zn/other ferrites have also been usually limited in range of room temperature to as low as 4 K [30]. But the most technological uses of these ferrites are either at or above room temperature. The hysteresis loop, magnetization as well as magnetic moment are strongly dependent on operating temperature. The room temperature measurement of magnetic properties cannot provide exact behavior of magnetic material at higher temperature since the temperature dependent magnetic behavior is highly non-linear. Therefore, to design a magnetic device using soft ferrites, thermal effects on magnetic properties need to be considered [31-32]. This prospect also motivates the researchers and some reports on the high temperature magnetic properties also available [30, 33-34]. Furthermore, the study of dc electrical resistivity provides important information regarding suitability of materials in high frequency application. In addition, determination of band gap of semiconducting materials is very important for their practical application. Therefore, the present paper reports the investigation of mechanical properties, high temperature magnetic properties along with FTIR spectra and UV-Vis measurement of Mg-Zn ferrites.

Mg-Zn ferrites have been subjected by a large number of research group to modify the different properties using different synthesis routes and conditions [12, 14, 35-49], where some of them are dealt with tailoring electrical properties and/or magnetic properties of Mg-Zn ferrites. The motivation of our present research is to enhance the electrical resistivity of the Mg-Zn ferrites by $Y^{3+}$ substitution for $Fe^{3+}$ with favorable magnetic properties that is one of the significant characteristics of ferrites for applications in high frequency devices. Melagiriyappa *et al.* [35] have been studied the electrical properties of Sm substituted Mg-Zn ferrites where increased of electrical resistivity has been reported. The initial permeability and saturation magnetization ($M_s$) are focus and the variation is also being explained with the variation of Cu contents as well as density of the samples [36]. Mn substituted Mg-Zn ferrites have been investigated by Mohseni *et* *al.* [37] in which the structural, vibrational, micro-structural and magnetic properties are considered. The increase in $M_s$ and decrease in coercivity due to Mn substitution is reported. The effect of $Cr^{3+}$ substitution in the Mg–Zn ferrite has been studied by Haralkar *et al.* [42] and Xia *et al.* [43]; they have reported on the decrease of $M_s$ with Cr contents. Mukhtar *et al.* [48] have investigated the Pr substituted Mg-Zn ferrites where the samples are characterized by XRD, FTIR spectroscope, SEM, EDX and VSM measurements. The decrease in $M_s$ and increase in coercive field ($H_c$) has been reported with increase in Pr contents. The improvement in Magnetic properties of Mg-Zn ferrites by Co substitution has also been reported [45]. Recently, the effect of Y substitution on the electrical resistivity and magnetic properties of different ferrites has been reported [34, 50-56]. But to the best of our knowledge, Y is not considered as a substituent in the Mg-Zn ferrites yet. We have studied the effect of Y substitution on structural properties, morphological study, ac conductivity, dielectric properties, permeability and room temperature magnetic properties of Mg-Zn ferrites and results are available elsewhere [57].

Therefore, in this article, FTIR spectroscopy, mechanical properties, dc resistivity, UV-Vis spectroscopy and temperature dependent magnetic properties of Y-substituted Mg-Zn ferrites prepared from nano-sized raw materials have been presented for the first time.

## 2. Materials and methods

A series of $Mg_{0.5}Zn_{0.5}Y_xFe_{2-x}O_4$ ($0 \le x \le 0.05$) ferrites, in the steps of 0.01, have been synthesized using analytical grade nano-sized materials by conventional ceramic technique. The use of nano powders as raw materials is inspired by successful completion some other projects that can be found elsewhere [58-61]. The detail of the sample preparation is presented elsewhere [57]. The sample preparation method comprises the following operations as shown in Fig. 1. The X-ray diffraction (XRD) spectra of the prepared samples has been recorded by a PHILIPS X'Pert PRO X-ray diffractometer in the range $2\theta$ = 15–70$^0$ using Cu $K_\alpha$ radiation ($\lambda$ = 1.54 Å). The microstructure images have been taken by the Field Emission Scanning Electron Microscope (FESEM) (model: JEOL JSM-7600F) equipped with an energy dispersive spectrometer (EDS). A PerkinElmer FTIR spectrometer has been used to compute the Fourier Transform Infrared (FTIR) spectra using the attenuated total reflection (ATR) technique in the range 400–1000 cm$^{-1}$. Elastic properties were calculated using FTIR data along with structural parameters. For dc electrical resistivity measurements, the as-prepared powders were pressed to pellet (diameter 8.4 mm, thickness 2.4 mm) and sintered. Conductive silver paint was applied and two probe method was used to measure the dc electrical resistivity. Temperature dependent M vs H curves and magnetic moment measurement were done by a Quantum Design PPMS VSM magnetometer.

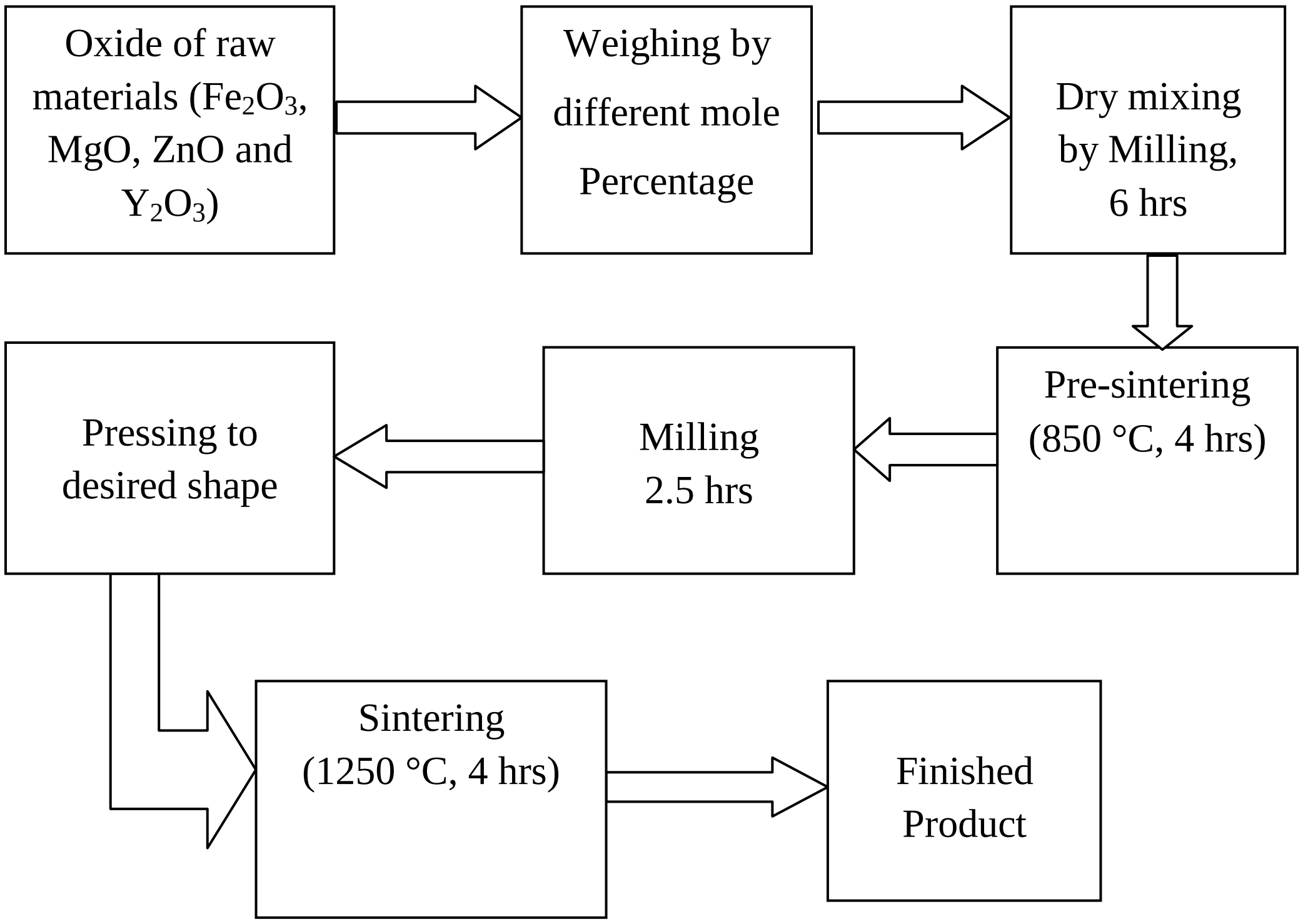

**Fig. 1.** Flow chart of sample preparation method.

## 3. Results and discussion

### *3.1 Phase analysis*

Fig. 2 depicts the XRD patterns of $Mg_{0.5}Zn_{0.5}Y_xFe_{2-x}O_4 (0 \leq x \leq 0.05)$ ferrites. The peaks are indexed for different atomic planes and compared with reported peaks of different ferrites [34, 50-56]. The XRD patterns reveal synthesis of spinel ferrites in which the single phase noted up to *x*=0.03. A secondary phase is appeared at 2θ=33.2° for higher Y contents. This extra phase (marked as red circle) results from $YFeO_3$ because of highly reactive tendency of $Fe^{3+}$ ions with $Y^{3+}$ ions [62]. This type of secondary phase is also shown in case of Y substitution in other ferrites [55-56]. The structural parameters have been calculated using the XRD data and can be found elsewhere [57].

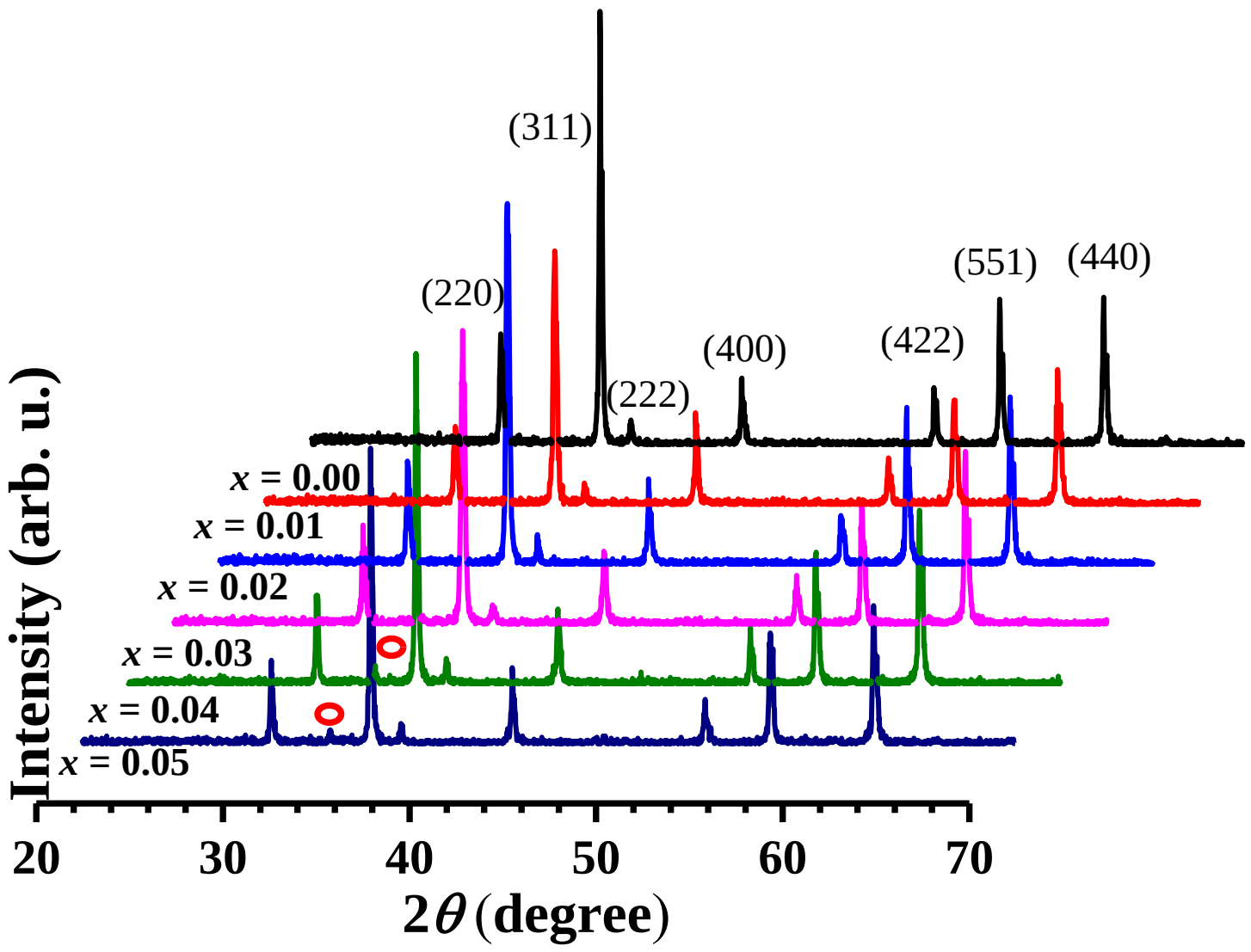

**Fig. 2.** The XRD patterns of $Mg_{0.5}Zn_{0.5}Y_xFe_{2-x}O_4 (0 \leq x \leq 0.05)$.

*3.2 Microstructure study*

The physical properties, especially the magnetic properties, are sensitive to the microstructure of ferrites and by manipulation of microstructure the properties can also be tailored. Fig. 3 illustrates the FESEM images of $Mg_{0.5}Zn_{0.5}Y_xFe_{2-x}O_4$ ($0 \leq x \leq 0.05$) where considerable modification in microstructure is observed due to the Y substitution. The EDS spectra for each composition are presented in Fig. 4(a-f). The qualitative analysis of EDS spectra endorses the existence of Mg, Zn, Fe, Y and O ions and nonattendance of any undesired element. The quantitative analysis of EDS spectra provides the ratio of metal cations and anions present in the samples which are also calculated [Table 1] and found in good agreement with other reported results [14].

**Table 1** Cations-anions % (At) and cations-anions ratio of $Mg_{0.5}Zn_{0.5}Y_xFe_{2-x}O_4$ ($0 \leq x \leq 0.05$).

| Y contents (*x*) | Grain size (μm) | cations (At %) | anions (At %) | Ratio of cation: anion |
|---|---|---|---|---|
| 0.00 | 1.22 | 40.03 | 59.97 | 2.802:4.197 |
| 0.01 | 3.65 | 40.68 | 59.32 | 2.847:4.152 |
| 0.02 | 3.75 | 40.29 | 59.71 | 2.820:4.179 |
| 0.03 | 1.28 | 42.09 | 57.91 | 2.946:4.053 |
| 0.04 | 1.07 | 47.94 | 53.94 | 3.355:3.775 |
| 0.05 | 1.87 | 38.40 | 61.60 | 2.688:4.312 |

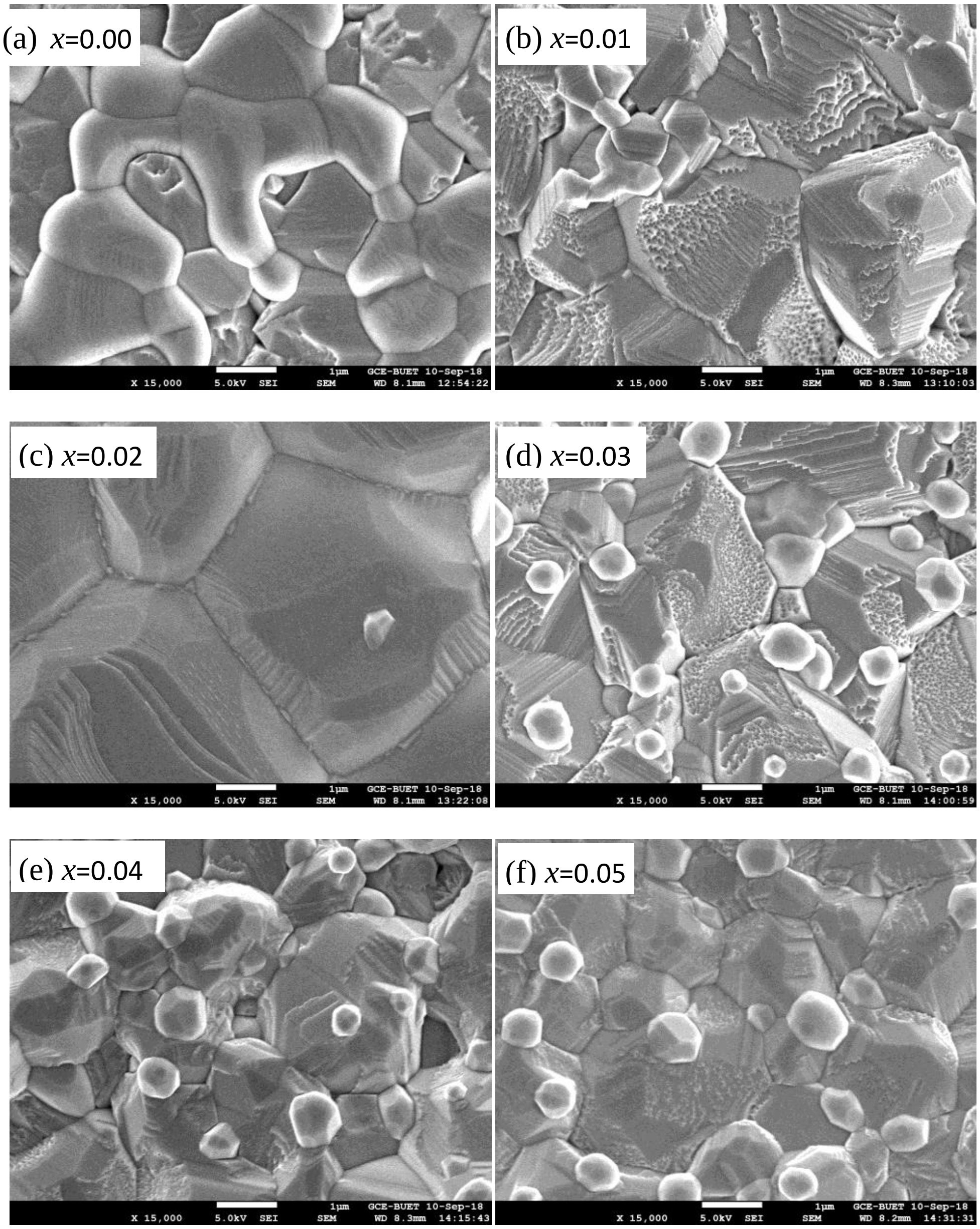

**Fig. 3**. FESEM images of $Mg_{0.5}Zn_{0.5}Y_xFe_{2-x}O_4$ ($0 \leq x \leq 0.05$) [57].

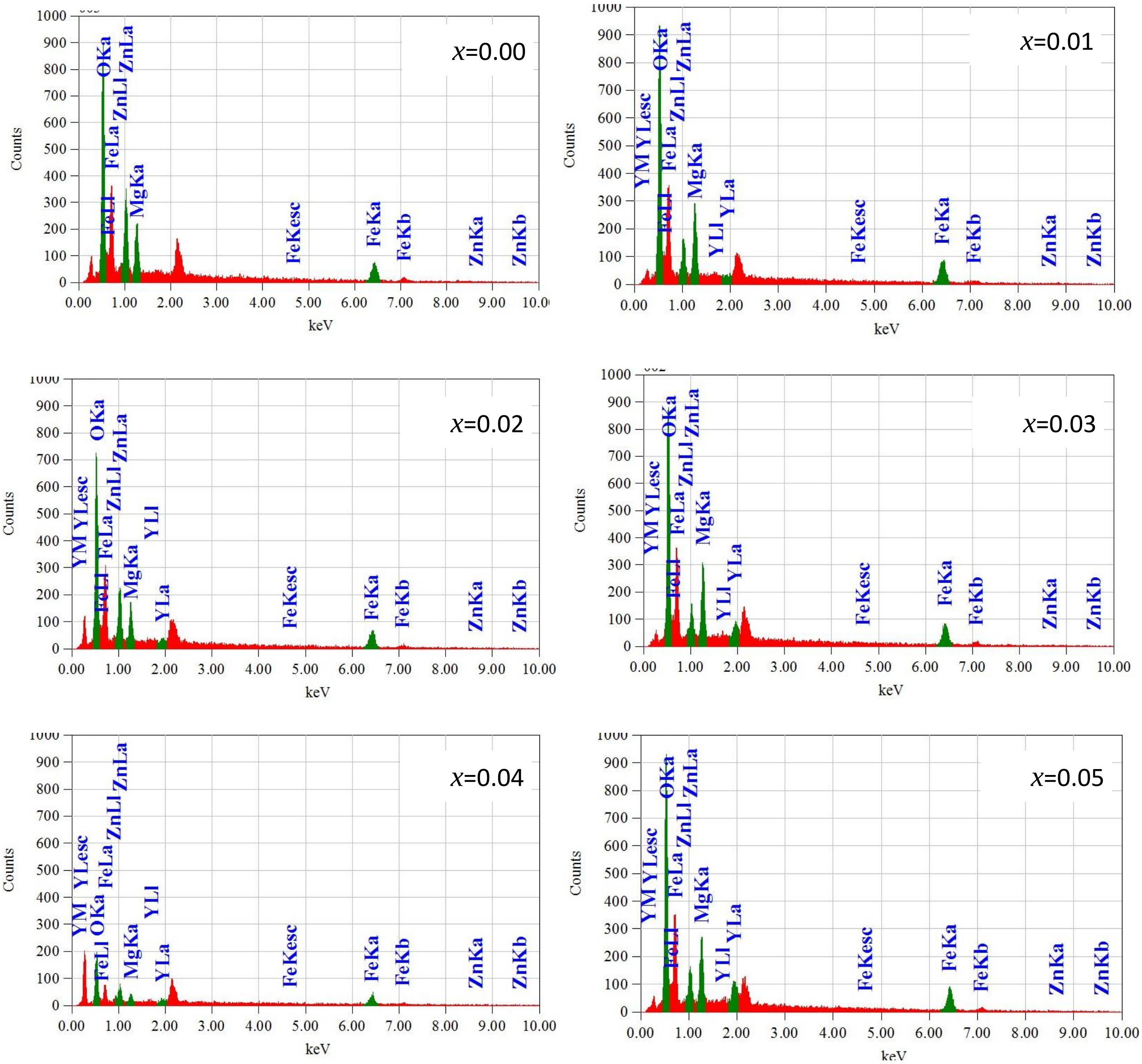

**Fig. 4**. EDS spectra of $Mg_{0.5}Zn_{0.5}Y_xFe_{2-x}O_4$ ($0 \leq x \leq 0.05$).

### 3.3. *FTIR spectroscopy*

FTIR spectra of ferrites generally exhibit two major peaks for the M-O vibrations at *A*- and *B*-sub-lattice [63]. Fig. 5 illustrates the FTIR spectra of the considered ferrite compositions in which two major peaks are noted within the expected range of frequency: 365 to 1100 cm$^{-1}$. Table 2 shows the peaks points for the studied compositions, the completion of solid state reaction has been confirmed from the position of the band within the expected range of frequency for ferrites with spinel structure [64]. The bands at higher frequency ($\nu_1$ = 560 cm$^{-1}$) correspond to the stretching vibrations of the Metal–Oxygen bunch in the *A*-sites and at lower frequency ($\nu_2$ = 396-402 $cm^{-1}$) are allocated for the stretching vibration of the Metal–Oxygen bunch in the *B*-sites, respectively [65]. The band frequency $\nu_1$ remains constant for all composition. The band frequency $\nu_2$ slightly shifts to lower frequency side for $x$ = 0.04 and higher frequency side at $x$ = 0.05, this shift in peak position is might be related to the appearance of impurity phase. These measurements (XRD, FESEM and FTIR) confirmed the successful incorporation of Y into Mg-Zn ferrites.

**Table 2**

FTIR absorption bands for $Mg_{0.5}Zn_{0.5}Y_xFe_{2-x}O_4$.

| Y ($x$) | $\nu_1$ ($cm^{-1}$) | $\nu_2$ ($cm^{-1}$) | FcT×$10^5$(dynes/cm) | FcO×$10^5$(dynes/cm) | $K_{av}$×$10^5$ (dynes/cm) |
|---|---|---|---|---|---|
| 0 | 560 | 400 | 2.29 | 1.17 | 1.73 |
| 0.01 | 560 | 400 | 2.29 | 1.17 | 1.73 |
| 0.02 | 560 | 400 | 2.29 | 1.17 | 1.73 |
| 0.03 | 560 | 400 | 2.29 | 1.17 | 1.73 |
| 0.04 | 560 | 396 | 2.29 | 1.14 | 1.72 |
| 0.05 | 560 | 402 | 2.29 | 1.18 | 1.74 |

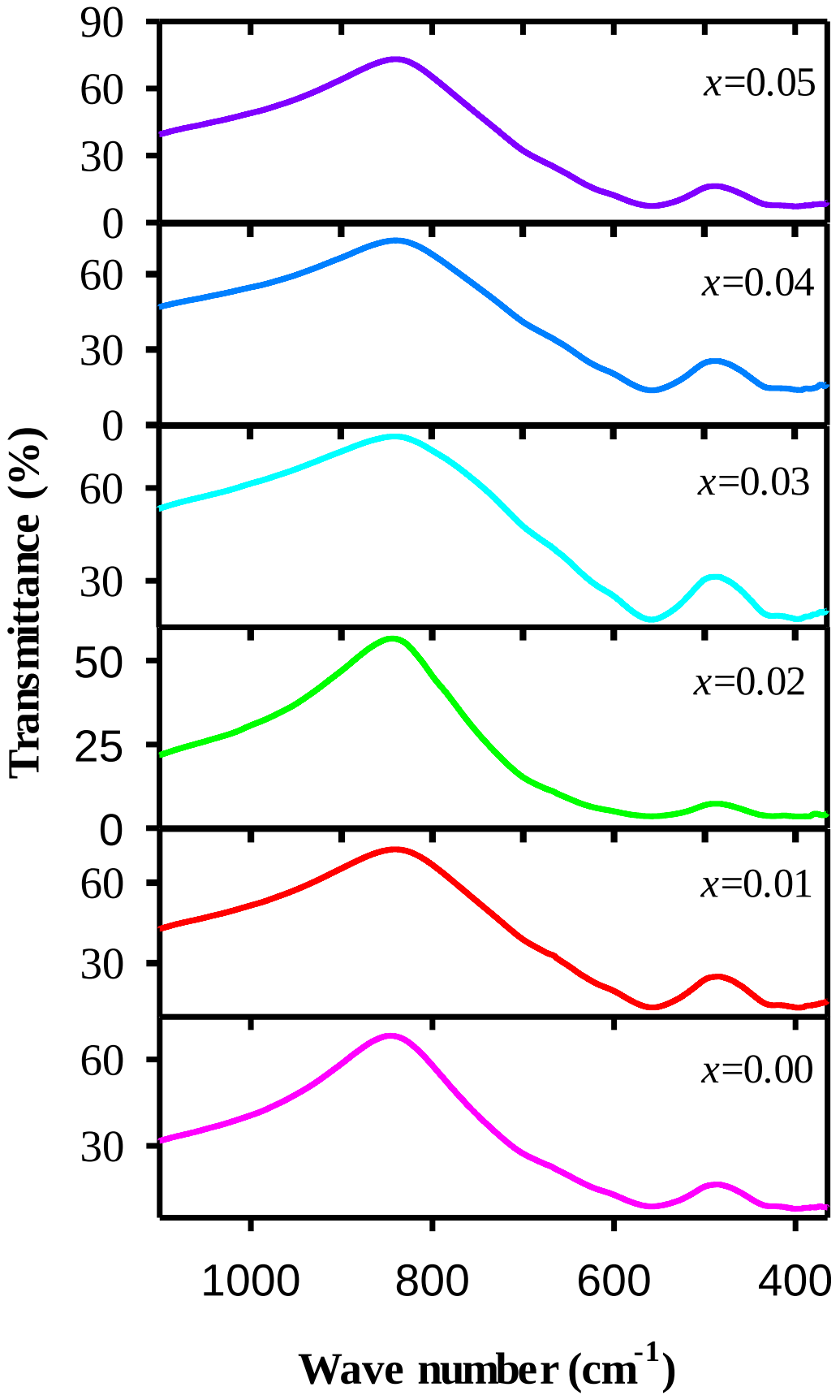

**Fig. 5.** FTIR spectra of $Mg_{0.5}Zn_{0.5}Y_xFe_{2-x}O_4 (0 \leq x \leq 0.05)$.

*3.4 Mechanical properties*

The stability and bonding characteristics among the adjacent atomic plane can be understood from the stiffness constants. For cubic crystal system, there are three stiffness constants: $C_{11}$, $C_{12}$ and $C_{44}$. The FTIR spectroscopy data can be used to determine the stiffness constant, elastic constants and Debye temperature etc. [66, 67]. The equations used to calculate $C_{11}$ is given by [68] $C_{11} = {K_{av}}/{a}$, where $K_{av}$ is the average of tetrahedral and octahedral force constant calculated [Table 2] from FTIR spectra using the following equation [69]:$F_c = 4\pi^2 c^2 \nu^2 m$, where, $c$, υ and $m$ are the speed of light in free space, vibrational frequency and the reduced mass of $Fe^{3+}$ and $O^{2-}$ ions, respectively. The equivalent reduced mass is about $2.061\times10^{-23}$ g and $a$ is the lattice constant. The relation [68]: $C_{12} = \frac{\sigma C_{11}}{1-\sigma}$ is used to estimate the $C_{12}$, where σ is the Poisson's ratio [70]: $\sigma = 0.324 \times [1 - (1.043 \times P)]$, P is the pore fraction. The $C_{44}$ is calculated from the relation: $C_{44} = \frac{C_{11}-C_{12}}{2}$. The obtained values of $C_{ij}$ [Table 3] confirmed that the ferrites compositions under consideration satisfy the Born criteria for mechanical stability [71, 72]: $C_{11} - C_{12} > 0$; $C_{11} + 2C_{12} > 0$; $C_{44} > 0$. Therefore, the synthesized compositions are mechanically stable. The stiffness constant $C_{11}$ represents the elasticity of length and provides the stiffness of materials along the (100) direction while the constants $C_{12}$ and $C_{44}$ born the elasticity in shape. The variation in stiffness constants with Y contents is shown in Fig. 6 (a).

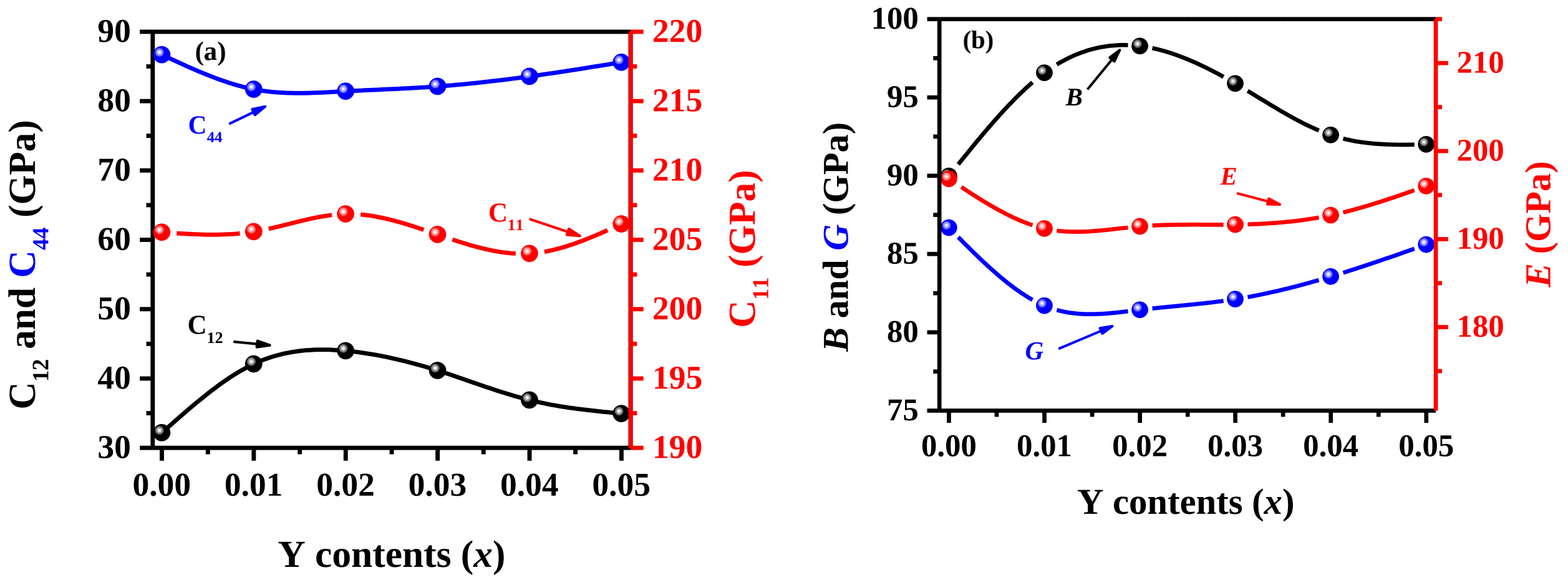

**Fig. 6.** The variation of (a) stiffness constants ($C_{ij}$) and elastic moduli with Y contents.

The values of $C_{11}$ and $C_{12}$ increase up to $x$ = 0.02 and then $C_{12}$ gradually decreases while $C_{11}$ decreases up to $x$ = 0.04 and increases further. The improvement in $C_{11}$ and $C_{12}$ endorse the good solubility of Y ions into $B$-sites and strengthen of the inter-atomic bonding. While the decline trend in $C_{11}$ and $C_{12}$ curves indicates the decrease in crystallization process. Microstructure images [Fig. 3] also ratify the crystallization of prepared samples. In this case, the variation in $C_{11}$ and $C_{12}$ and bulk density [Table 3] may be attributed from the variation of grains size that can be observed from Fig. 3. [73]. The equations [70]:$E = \frac{(C_{11}-C_{12})(C_{11}+2C_{12})}{(C_{11}+C_{12})}$, $B = \frac{1(C_{11}+2C_{12})}{3}$, $G = \frac{E}{2(\sigma+1)}$ are used to calculate the Young's modulus (*E*), Bulk modulus (*B)* and shear modulus (*G*) and presented in Table 3. Fig. 6 (b) represents the variation in different elastic moduli [*B*, *G* and *E*] with Y contents (*x*). The bulk modulus provides the value of required resistance for volume deformation, an increase is observed up to *x* = 0.02 due to improvement in microstructure and then decreases gradually. This variation might be related to the variation in bulk density as well as the grains size (*x*) [73]. The shear and Young's modulus are noted to decrease for *x* = 0.01 and then increase gradually with the variation of *x*. The value of *E* provides the stiffness of solids where the high value of *E* indicates the stiffer solids. The values of *E* and *G* of substituted compositions are lower than that of the parent one [*x*= 0.00]. The change in inter-atomic bonding is responsible for the variation in elastic moduli [26]. Although, the variation of elastic moduli [Fig. 6] is shown as a function of Y contents in magnified scale, but if we have a look on the values presented in Table 3, an insignificant variation is observed. Table 3 shows the Poisson's ratio [$\sigma$] for all compositions. The Poisson's ratio is extremely important in materials application point of view which determines the ductile behavior or brittleness of materials with a decisive factor of 0.26 [74, 75]. A value of σ less than 0.26 indicates the brittleness of materials otherwise ductile behavior of materials. Like other ceramics materials, brittleness of studied compositions is confirmed and the values of $\sigma$ for all compositions lie in the expected range from -1 to 0.5, confirming excellent elastic behavior with the theory of elasticity [76]. The anisotropy constant *A* ($A = 2C_{44}/(C_{11}-C_{12})$) [77-78] also calculated [Table 3] and found to have the value of 1 (one) for all composition, indicating isotropic nature of cubic ferrites compositions.

The Debye temperature ($\Theta_D$), the most fundamental property of solids, used to resolve many of characteristics thermal properties; determine the lower temperature region and higher temperature region for solids. Moreover, the $\Theta_D$ also provides the information regarding bonding strength. The higher value of $\Theta_D$ is associated with stronger bonds in solids. The average sound velocity can be used to calculate $\Theta_D$ by the following equation [79]: $\Theta_D = {}^{h}/_{k_B}\,[({}^{3n}/_{4\pi})N_A\rho/M]^{1/3}\,v_m$, where the symbols signify the usual meanings [75]. The average sound velocity is calculated by the equation [79]: $v_m = \left[1/3\left(1/v_l^3 + 2/v_s^3\right)\right]^{-1/3}$, where $v_l$ [$v_l = [(3B+4G)/3\rho]^{1/2}$] and $v_s$ [$v_s = [G/\rho]^{1/2}$] are the longitudinal and transverse sound velocities in an isotropic material.

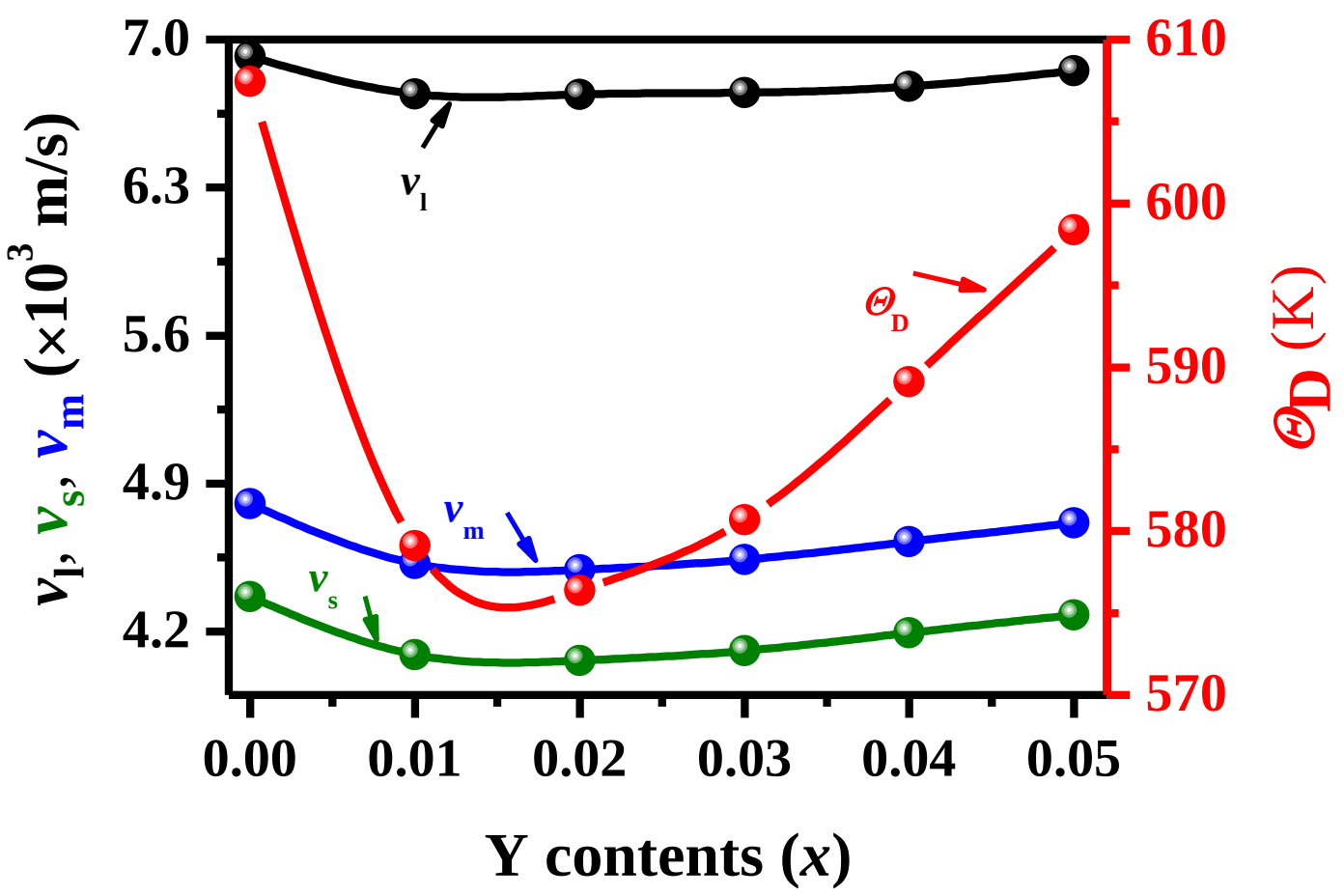

**Fig. 7.** The variation of sound velocities and Debye temperature with Y contents in the composition of $Mg_{0.5}Zn_{0.5}Y_xFe_{2-x}O_4$ ($0 \leq x \leq 0.05$).

**Table 3**

The lattice constant ($a$), bulk density ($\rho_b$), porosity ($P$), average force constant ($K_{av}$), stiffness constant ($C_{ij}$) elastic moduli ($E$, $B$, $G$), Poisson's ratio ($\sigma$), anisotropic constant ($A$), sound velocities ($v_l$, $v_s$, $v_m$) and Debye temperature ($\Theta_D$).

| Parameters | $x$ =0.00 | $x$ =0.01 | $x$ =0.02 | $x$ =0.03 | $x$ =0.04 | $x$ =0.05 |
|---|---|---|---|---|---|---|
| $a$ (Å) | 8.4309 | 8.4327 | 8.4366 | 8.4350 | 8.4269 | 8.4288 |
| $\rho_b$(kg/m$^3$) | 4006 | 4172 | 4197 | 4166 | 4119 | 4084 |
| $P$ (%) | 18.08 | 14.76 | 14.25 | 15.06 | 16.38 | 17.17 |
| $K_{av.}$ (N/m) | 173.40 | 173.40 | 174.58 | 173.40 | 172.24 | 173.99 |
| $C_{11}$ (GPa) | 205.55 | 205.59 | 206.86 | 205.38 | 204.01 | 206.14 |
| $C_{12}$ (GPa) | 32.19 | 42.10 | 43.99 | 41.15 | 36.89 | 34.93 |
| $C_{44}$ (GPa) | 86.68 | 81.70 | 81.43 | 82.11 | 83.56 | 85.60 |
| $E$ (GPa) | 196.83 | 191.18 | 191.43 | 191.64 | 192.71 | 196.01 |
| $B$ (GPa) | 89.98 | 96.57 | 98.28 | 95.89 | 92.60 | 92.00 |
| $G$ (GPa) | 86.68 | 81.70 | 81.43 | 82.11 | 83.56 | 85.60 |
| $\sigma$ | 0.13 | 0.17 | 0.17 | 0.16 | 0.15 | 0.14 |
| $A$ | 1.0 | 1.0 | 1.0 | 1.0 | 1.0 | 1.0 |
| $v_1$ (m/s) | 6918.9 | 6742.7 | 6740.1 | 6748.3 | 6777.8 | 6851.1 |
| $v_s$ (m/s) | 4365.4 | 4091.0 | 4063.0 | 4109.9 | 4194.3 | 4278.7 |
| $v_m$ (m/s) | 4803.9 | 4520.7 | 4492.6 | 4539.8 | 4625.3 | 4713.8 |
| $\Theta_D$ (K) | 607.43 | 579.12 | 576.38 | 580.70 | 589.12 | 598.39 |

The calculated sound velocities mentioned are presented in Table 3 along with the calculated $\Theta_D$. It is observed [Table 3] that the transverse sound velocity is lower than the longitudinal sound velocity. Energy is transferred from particle to particle by vibration of particle during propagation of wave through a medium. During the propagation of transverse wave, the vibration of particle is at right angles with the direction of wave transmission, caused a higher energy requirement to vibrate the adjacent particle. As a result the energy of the wave reduced and hence the velocity of transverse wave is lower than the longitudinal wave and is roughly half of longitudinal wave velocity [80]. Fig. 7 illustrates the variation of $\Theta_D$ with Y contents in which it seen that the $\Theta_D$ decreases with Y contents up to $x$ = 0.02 and then increases with Y contents. The similarity in the variation of sound velocities and $\Theta_D$ is observed similar to other ferrites [26, 73, 76]. The variation of $\Theta_D$ roughly follows the variation of rigidity modulus [76] as can be shown in Fig. 6 (b).

3.5 *dc resistivity*

High resistivity of ferrites is essential for most electronic application and therefore, the study of dc electrical resistivity is important for prediction of their application in electronic and high frequency devices. The room temperature dc electrical resistivity ($\rho_{RT}$) of $Mg_{0.5}Zn_{0.5}Y_xFe_{2-x}O_4$ ($0 \leq x \leq 0.05$) is estimated by the following equation: $\rho_{RT} = \frac{RA}{L}$, where, $R$, $A$ *and* $L$ is resistance, area and thickness, respectively of the pellet. The value of $\rho_{RT}$ is noted to increased due to Y substitution unlike for $x$ = 0.01 (Fig. 8 (a)), can be explained by Verwey mechanism based on electrons exchange involving the ions of the element with different valence states positioned octahedral (*B*-site) [81-82]. In the present case, the $Fe^{2+}$ and $Fe^{3+}$ is responsible for the conduction because of the hopping tendency between them [83, 84] and the $Fe^{2+}$ ions are the product of sintering process [42]. The value of $\rho_{RT}$ is found to be $0.45 \times 10^5$ Ω-cm for x = 0 and $0.40 \times 10^5$ Ω-cm, $1.57 \times 10^6$ Ω-cm, $1.78 \times 10^6$ Ω-cm, $2.19 \times 10^6$ Ω-cm and $1.43 \times 10^6$ Ω-cm for $x$ = 0.01, 0.02, 0.03, 0.04 and 0.05, respectively. Fig. 8 (a) clearly indicates the increase in resistivity for substituted composition (except for $x$ = 0.01) where the resistivity decreases slightly. The ac resistivity was found to decrease for $x$ = 0.01 [57]. The Y ions replace $Fe^{3+}$ ions at *B*-sites. [50], simultaneously small fraction of Zn ions forfeit balanced by forcing some $Fe^{3+}$ to move at the *A*-sites. The Fe ions have higher electronic valence than Zn ions; hence the charge carriers (here, metallic vacancy) have been increased by balancing the electrical charge, results decrease in resistivity. Two mechanisms might be involved in increasing resistivity, firstly, the enhancement in resistivity due to Y substitution is attributed by forming the stable bonds (electric) of $Y^{3+}$ and $Fe^{2+}$, causes localization of $Fe^{2+}$, limiting the Verwey mechanism [81], hence upturn the resistivity. Secondly, the preference of $Y^{3+}$ ions to reside in the *B*-sites, lead to a replacement of $Fe^{3+}$ ions at *B*-sites, results a reduction of $Fe^{2+}$ formation. Since, $Y^{3+}$ ions do not take place in the conduction due to its stable oxidation state and electrons hopping takes place between $Fe^{3+}$ and $Fe^{2+}$, therefore, the degree of electrons change ($Fe^{2+} \leftrightarrow Fe^{3+}$) is reduced and consequently the resistivity is increased. The resistivity of materials greatly depends on its microstructure: the porosity and the distribution of grains. The lowering of resistivity for $x = 0.05$ contents may be explained using the microstructure [Fig. 3(f)] of substituted compositions and the improved connectivity between the grains [85]. From Fig. 3 (d-e), it is seen that the grains distribution are more homogeneous for $x = 0.05$ than that for $x = 0.03$ and 0.04; and expected to have the improved connectivity between the grains. Therefore, a slight decrease in resistivity is also expected.

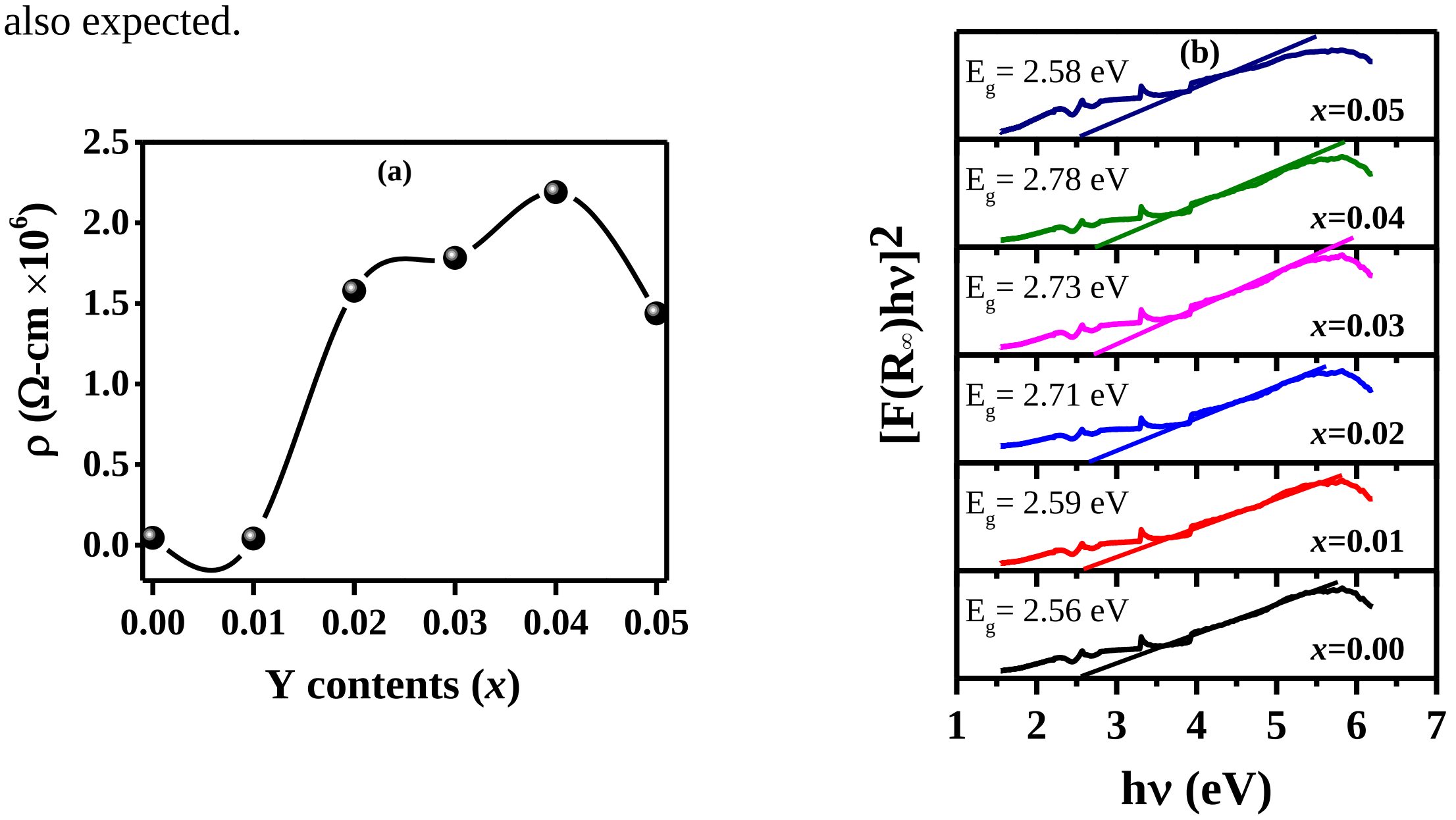

**Fig. 8.** (a) variation dc resistivity with Y contents, and (b) plot of $[F(R_\infty)h\nu]^2$ against $h\nu$ of $Mg_{0.5}Zn_{0.5}Y_xFe_{2-x}O_4$ ($0 \leq x \leq 0.05$) ferrites.

3.6 *Diffuse reflectance results*

To know the optical properties of Y-substituted Mg-Zn ferrites, the UV–Vis diffuse reflectance study has been performed. The optical reflectance data can be used to calculate the band gap using the equation $F(R_\infty) = \frac{A(h\nu - E_2)^n}{h\nu}$, where, $F(R_\infty)=(1-R)^2/2R_\infty$, is the Kubelka-Munk Function [86]. Therefore, the plot of $[F(R_\infty)h\nu]^2$ against photon energy (hν) gives the way of calculating the energy band gap $E_g$ (eV), by extrapolating the linear part to the x-axis (energy axis). The band gap is found to be increased due to the Y substitution that is consistent with the Y-substituted Co ferrites [53]. The variation in band gap indicates the change in structural parameters attributed from the change in Y contents [53]. The band gap enhancement is associated with the variation of dc resistivity [Fig. 8 (b)].

### 3.7 *Magnetic properties*

#### 3.7.1 *High temperature magnetic properties mapping*

The magnetization vs applied magnetic field (upper part of hysteresis loops) at different measuring temperatures, as shown in Fig. 9 (a-f), exhibiting typical magnetic nature at elevated temperature, revealing the existence of magnetic ordered structure near $T_c$ (in section 3.7.3). Beyond $T_c$, destruction of magnetic order is also observed i.e., transition from ferrimagnetic (below $T_c$) to paramagnetic (above $T_c$) nature. It is [Fig. 9 (a-f) also expected that the coercivity ($H_c$) and remanent magnetization ($M_r$) decrease with increasing temperature and tend to be zero near $T_c$. It is very important that the ferrimagnetic ordered structured remains up to $T_c$ indicating the suitability of these materials to use them at higher temperature.

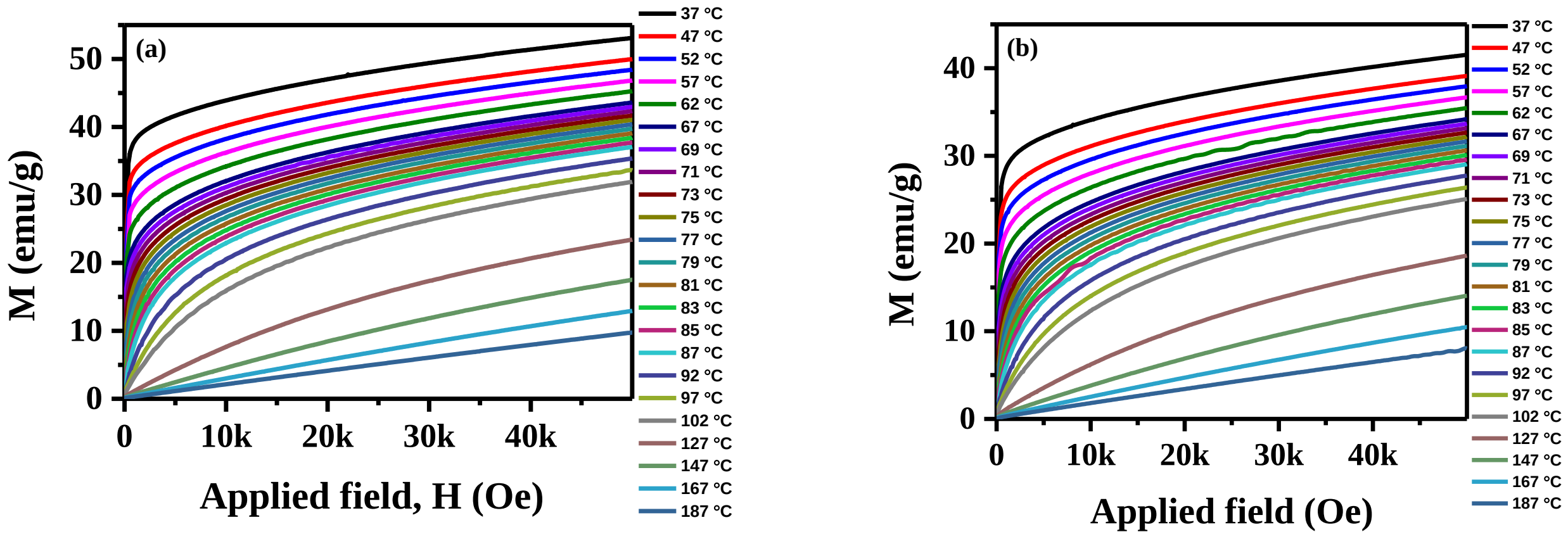

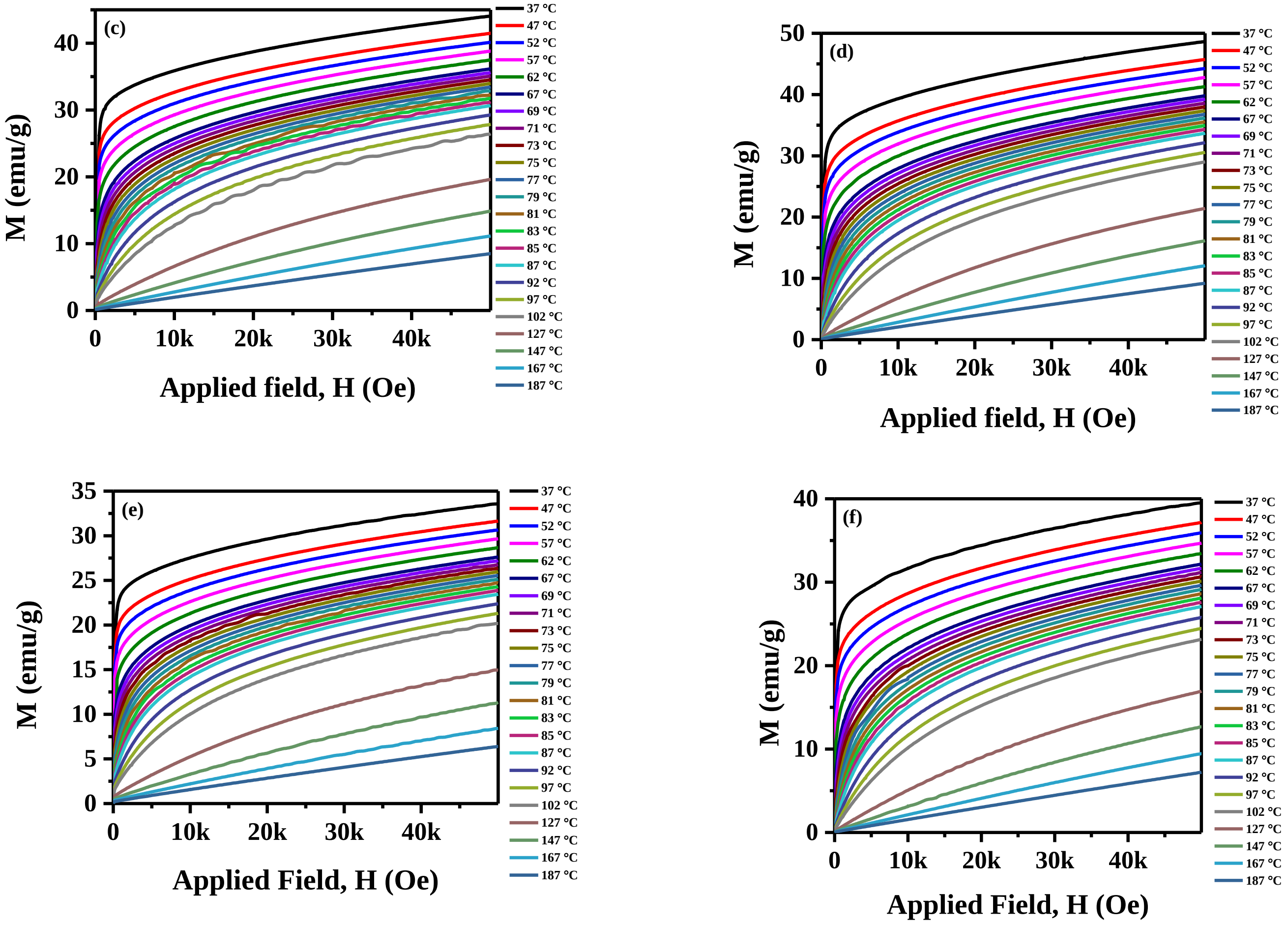

**Fig. 9 (a-f)** The M vs H plot of $Mg_{0.5}Zn_{0.5}Y_xFe_{2-x}O_4$ ($0 \leq x \leq 0.05$) measured at different temperatures.

The value of $M_s$ for each composition at different selected temperatures is obtained by the following: at first the $M_s$ vs 1/H curves are plotted (figures not shown here) and are extrapolated to zero i.e., 1/H = 0 [53]. The obtained $M_s$ values are plotted in Fig. 10 (a) for different Y contents at different temperature. The variation of $M_s$ with Y contents has been explained elsewhere [57]. A decrease in the $M_s$ with increasing temperature is observed due to more randomness of magnetic moments that attributed from the increase in thermal energy. A change in site occupancy of cation may also induce due to thermal fluctuation; results reduced $M_s$ at higher temperatures.

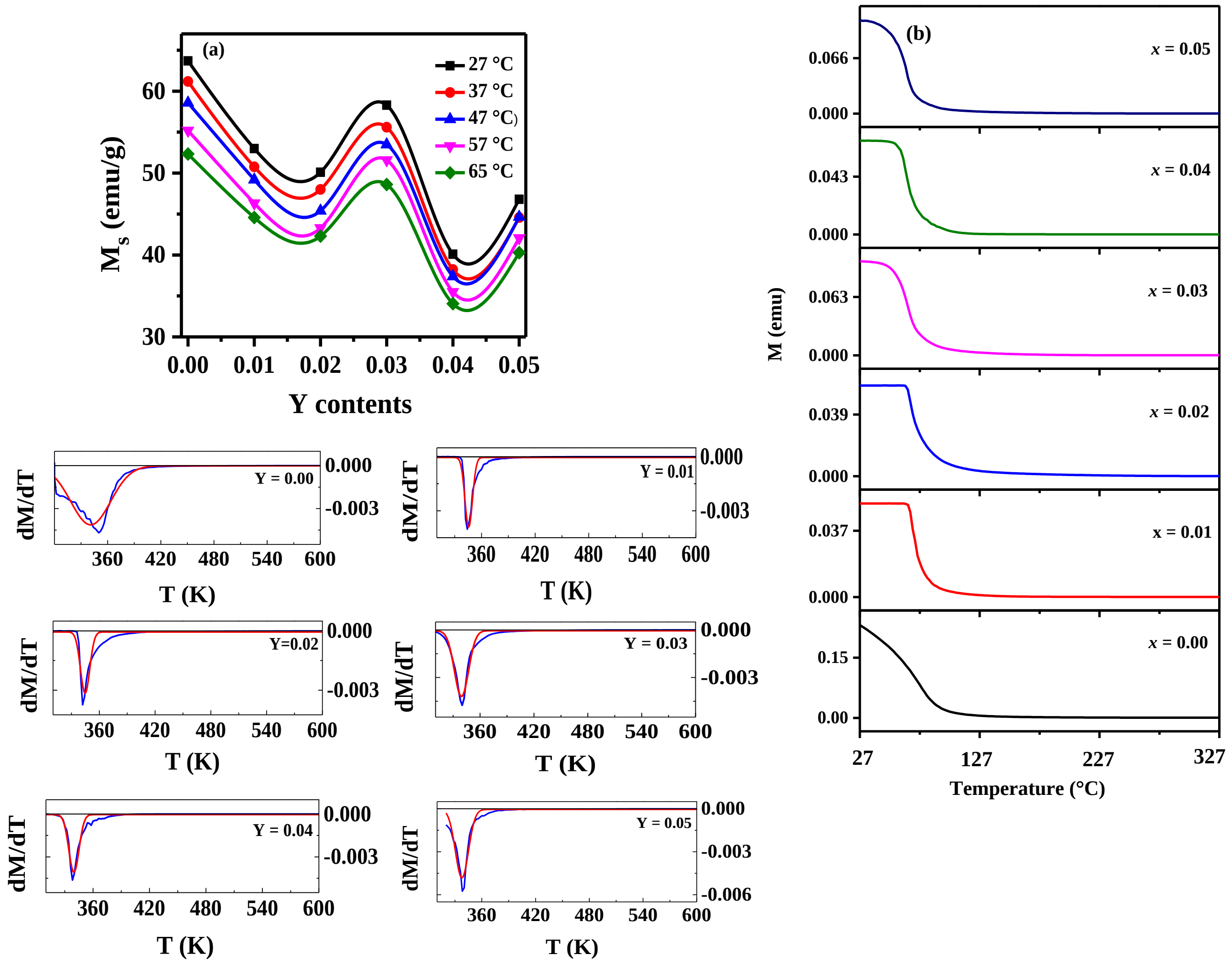

**Fig. 10.** Variation of (a) saturation magnetization at different measuring temperatures, (b) temperature dependent magnetic moment and (c) dM/dT vs temperature graph for different Y contents.

It is interesting to note that the width of the magnetic transition, inferred from the magnitude of the FWHM of the dM/dT vs T curve [Fig. 10 (c)] and the magnetic ordering temperature (taken at the peak of the dM/dT versus T graph) show features strongly dependent on the level of Y substitutions. For the Y substituted compounds [Fig. 10 (c)], the transition width does not show any systematic variation with Y content. The Y-free compound is characterized by a very large transition width. The transition temperature lies within a narrow range of 341 ± 4 K. This implies that the average mean-field exchange energy among the spins remains insensitive to the Y content. It should be noted that, substitutional disorder always introduces local strains inside the compounds. These local distortions and inhomogeneities lead to a distribution in the exchange energies. Within the mean-field approximation, the Curie temperature is directly proportional to the exchange energy [87]. Structure specific corrections often lower the theoretically predicted transition temperature, but this link between transition temperature and exchange energy remains valid. Within this framework, width of the magnetic transition is directly related to the distribution and homogeneity of atoms inside the solid. Therefore, different transition widths in the Y substituted compounds reflect the distribution and inhomogeneity of the exchange interactions inside these samples.

3.7.2 *Curie temperature*

The magnetic moment and initial permeability and hence the magnetic properties are very sensitive to temperature. Both exhibit an abrupt dropping near the transition point from ferrimagnetic to paramagnetic state. Therefore, the Curie temperature ($T_c$) has been measured by applying the two basic characteristics: (i) the samples are subjected to an applied magnetic field of 1000 Oe and the magnetic moment is calculated at different temperature. The heating rate is kept slow to avoid the nucleation and particle growth at elevated temperatures. The variation of magnetic moment with temperature is shown in Fig.10 (b) where the magnetic moment is found to decrease with increasing temperature as expected and tends to zero at different measuring temperature for different compositions. The $T_c$ value is then calculated from the first derivative of magnetic moment against temperature graph, Fig. 10 (c). Moreover, the $T_c$ is also calculated from the temperature dependent initial permeability, $\mu_i$ plot as shown in Fig.11 (a). A peak, also known as Hopkinson peak, is observed from the graph where the value of initial permeability dropped sharply. The characteristic temperature at which the peak observed is known as the $T_c$ where the spin magnetic moments are completely disordered and the conversion from ferrimagnetic to paramagnetic materials is taken place. The obtained $T_c$ is found to be decreased with the Y substation. The value of $T_c$ depends on the cation distribution on *A*- and *B*-site and the exchange coupling constant [53]. The coupling constant due to exchange interaction between ions in sub-lattice (*A* and *B*) is much stronger than the coupling constant due to exchange interaction among the ions in the same sub-lattice sites (*A/B*). The substitution of Y ions in the spinel structure leads to increase in the bond lengths and distance between ions presents at the *A*- and *B*–sites that results a decrease in exchange constants and hence the $T_c$ values [53]. The $T_c$ value for the $Mg_{0.5}Zn_{0.5}Fe_2O_4$ ($x$ = 0.00) is obtained from the M-T graph and $\mu'_i$-T graph and found to be 77°Cand 75°C, respectively which is higher than that of ($T_c$ ~47°C) reported by Khot *et al.* [88] and Hashim *et al.* [44] and lower than that of ($T_c$ ~97°C) obtained by Mazen *et al.* [89].

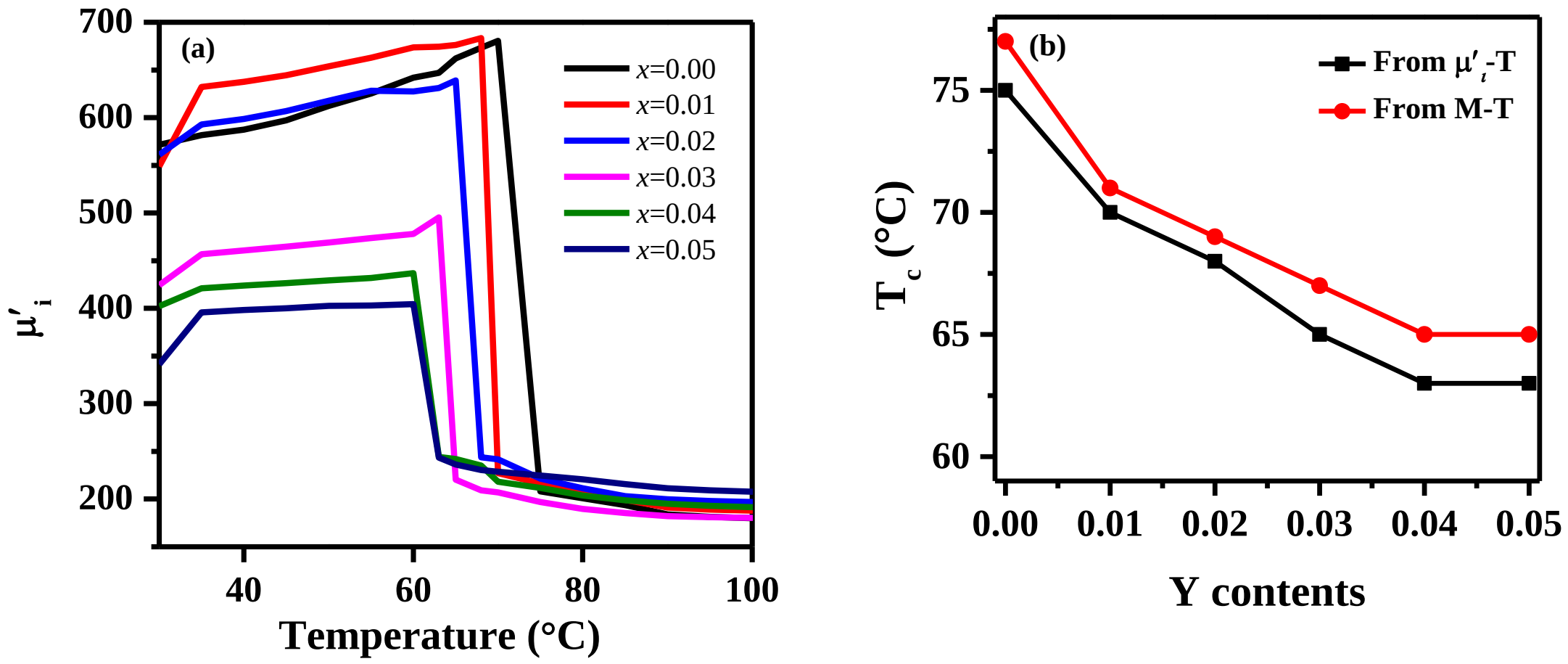

Fig. 11. (a) The temperature dependent initial permeability, and (b) the variation of Curie temperature with Y contents.

### 3.7.3 *Phase transition*

In section 3.7.1, the ferrimagnetic to paramagnetic phase transition has been observed and the transitions temperatures have also been obtained (section 3.7.2). The determination of the nature of phase transition (first or second order) is also significant. The Arrott plot ($M^2$ $(emu/g)^2$ against H/M (Oe.g/emu)) has been used to see the nature of phase transition. Banerjee [90] has proposed the criteria for first and second order transition. The positive slope of Arrott plot indicates the second order phase transition otherwise the transition is first order. The curves intercept at $M^2$ (Y-axis) indicate the presence of spontaneous magnetization while the curve passing through the origin measure the $T_c$ [91]. Fig. 12 (a-f) illustrates that the second order phase transition has been occurred in the prepared samples. It is also observed that the spontaneous magnetization is present up to ≤ 75 °C for $x$ = 0.00, ≤ 69 °C for $x$ = 0.01, ≤ 67 °C for $x$ = 0.02, ≤ 65 °C for $x$ = 0.03 and ≤ 63 °C for $x$ = 0.04 and 0.05 and thereafter the curves (after $T_c$ for each compositions) are straight upwards from H/M axis ($x$-axis) which indicates the paramagnetic behavior.

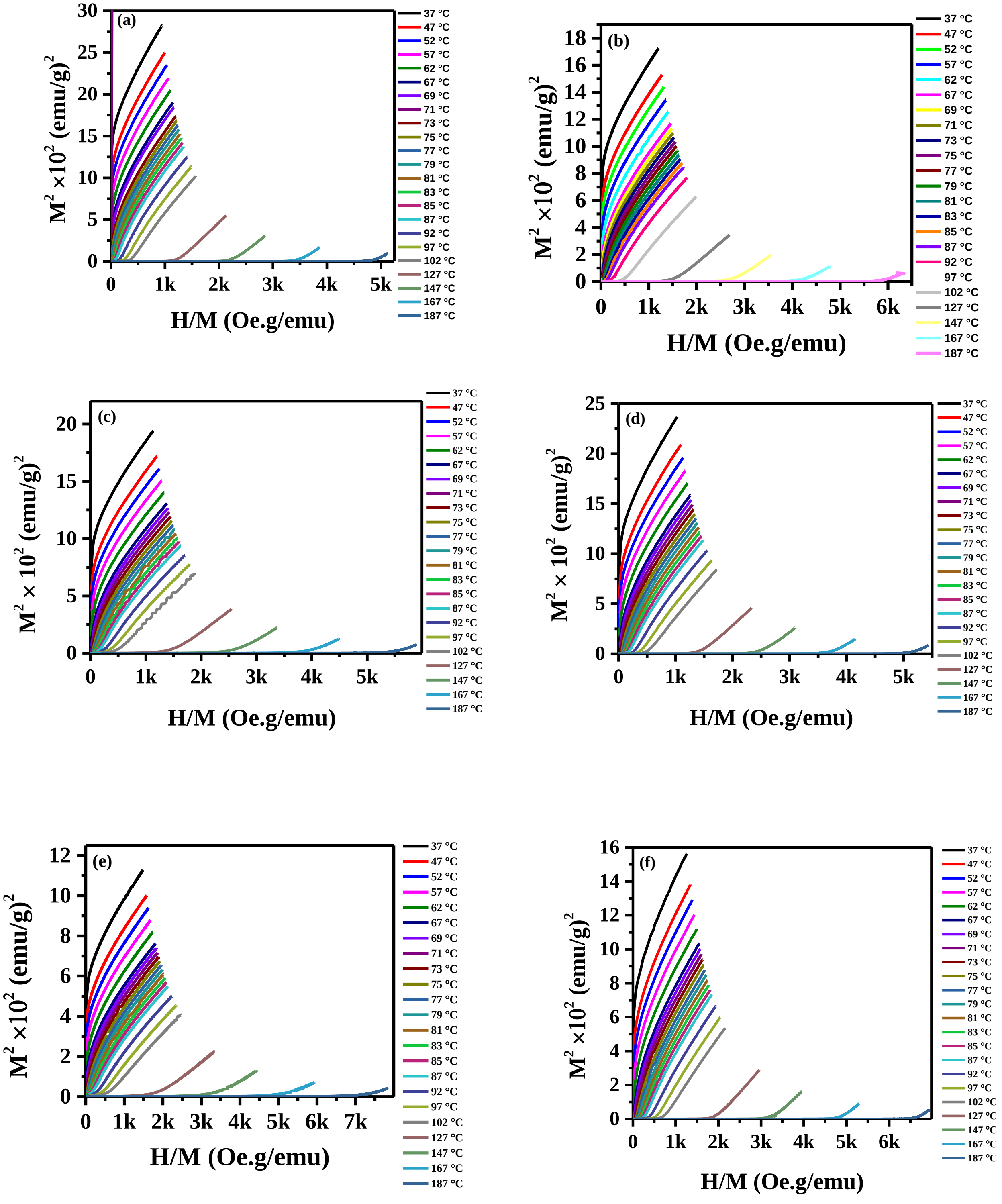

**Fig. 12**. The Arrott plot of $Mg_{0.5}Zn_{0.}5Y_{x}Fe_{2-x}O_4$ ($0\leq x \leq 0.05$) ferrites.

## 4. Conclusions

The Yttrium-substituted Mg-Zn [$Mg_{0.5}Zn_{0.5}Y_xFe_{2-x}O_4$ ($0 \leq x \leq 0.05$)] ferrites, with spinel structure as confirmed from the XRD and FTIR studies, have been synthesized using ceramic technique. The calculated stiffness constant revealed the mechanical stability of the compositions. The bulk modulus is increased for substituted compositions up to $x$ = 0.2 and decreased thereafter. The shear modulus and Young's modulus are found to be decreased at $x$ = 0.01 and then increased with further increase in Y contents. The values of Poisson's ratio (0.13 to 0.17) indicating the brittleness character of studied compositions. The value of anisotropic constants ($A$ = 1) be a sign of isotropic nature of studied spinel ferrites. The variation of $\Theta_D$ with increasing Y contents is associated with the variation in rigidity of the compositions. The dc resistivity increases from $0.45\times10^6$ Ω-cm (for $x$ = 0.00) to $1.57\times10^6$ Ω-cm, $1.78\times10^6$ Ω-cm, $2.19\times10^6$ Ω-cm and $1.43\times10^6$ Ω-cm for $x$ =0.02, 0.03, 0.04 and 0.05, respectively. The obtained optical energy band gap are found to be 2.56 eV (for $x$ = 0.00) to 2.59 eV, 2.71 eV, 2.73 eV, 2.78 eV and 2.58 eV for $x$ = 0.01, 0.02, 0.03, 0.04 and 0.05, respectively. Both the ferrimagnetic and paramagnetic nature have been confirmed from temperatures dependent magnetization curves. Very close values of $T_c$ obtained from the M-T and $\mu'_i$-T is noticed which is decreased due to the weakening of exchange coupling constant with Y substitution. The second order phase transition (ferrimagnetic to paramagnetic) has been occurred. The prepared ferrites are suitable for high frequency application as well as at temperature higher than room temperature (up to $T_c$).

## Acknowledgments

The authors are grateful to the Directorate of Research and Extension, Chittagong University of Engineering and Technology, Chattogram-4349, Bangladesh under grant number CUET DRE (CUET/DRE/2016-2017/PHY/005) for arranging the financial support for this work. We are thankful for the laboratory support of the Materials Science Division, Atomic Energy Center, Dhaka-1000, Bangladesh.